\documentstyle[epsf,12pt]{article}
\newcommand{\gsim}{\mbox{\raisebox{-.3em}{$\stackrel{>}{\sim}$}}}
\newcommand{\lsim}{\mbox{\raisebox{-.3em}{$\stackrel{<}{\sim}$}}}

\renewcommand{\cite}[1]{\ref{#1}}
\newcommand{\BBox}{\mbox{\raisebox{-0.2em}{\large$\Box$}}}
\newcommand{\half}{\frac{1}{2}}

\newcommand{\beq}{\begin{equation}}
\newcommand{\eeq}{\end{equation}}
\newcommand{\beqa}{\begin{eqnarray}}
\newcommand{\eeqa}{\end{eqnarray}}
\newcommand{\bpc}{\begin{picture}}
\newcommand{\epc}{\end{picture}}
\newcommand{\bfg}{\begin{figure}}
\newcommand{\efg}{\end{figure}}
\newcommand{\bcent}{\begin{center}}
\newcommand{\ecent}{\end{center}}

\newcommand{\reflef}{(\ref}
\newcommand{\nnb}{\nonumber}

\begin{document}
\baselineskip=0.8cm
\begin{center}
{\Large\bf Choosing a Conformal Frame in Scalar-Tensor Theories of
Gravity with a Cosmological Constant}\vspace{.3cm}\\
Yasunori Fujii$^{\dagger}$\\
Nihon Fukushi University, Handa, 475\ Japan\\
and\\
ICRR, University of Tokyo, Tanashi, Tokyo, 188\ Japan \vspace{.5cm}\\
\end{center}

\baselineskip=0.6cm

\begin{center}
{\large\bf Abstract}
\end{center}
Cosmological solutions of the Brans-Dicke theory with an added cosmological constant are investigated with an emphasis to select a
conformal frame in order to implement the scenario of a decaying cosmological constant, featuring an ever growing scalar field. We focus particularly on Jordan
frame, the original frame with the nonminimal coupling,
and conformally transformed Einstein frame without it. For the asymptotoic attractor solutions as well as the ``hesitation behavior," we find that none of
these conformal frames can be accepted as the basis of analyzing
primordial nucleosynthesis.  As a remedy, we propose to modify the prototype BD theory, by introducing a scale-invariant scalar-matter coupling, thus making Einstein frame acceptable.  The invariacne is broken as a quantum anomaly effect due to the non-gravitational interactions, entailing naturally the fifth force, characterized by a finite force-range  and WEP violation.  A tentative estimate shows that the theoretical prediction is roughly consistent with the observational upper bounds.  Further efforts to improve the experimental accuracy is strongly encouraged.\vspace{5cm} \\

\footnoterule
$\dagger$\ Electronic address: fujii@handy.n-fukushi.ac.jp
\newpage
\section{Introduction}

The cosmological constant is a two-step problem.  First, the observational {\em upper} bound to $\Lambda$ is more than 100 orders smaller than what is expected naturally from most of the models of unified theories.  Secondly, some of the recent cosmological findings seem to suggest strongly that there is a {\em lower} bound as well [\cite{ost}],  though it might be premature to draw a final conclusion.  To understand the first step of the problem, theoretical models of a ``decaying cosmological constant" have been proposed [\cite{dol},\cite{fn1}].  They are based on some versions of scalar-tensor theories of gravity essentially of the Brans-Dicke type [\cite{bd}].  Attempts toward the second step have also been made by extending the same type of theories [\cite{hwnt}].

As a generic aspect of the scalar-tensor theories, however, one faces an inherent question on how one can select a physical conformal frame out of two obvious alternatives, conveniently called J frame (for Jordan) and E frame (for Einstein), respectively.  The former is a conformal frame in which there  is a ``nonminimal coupling" that characterizes the Jordan-Brans-Dicke theory, but can be removed by a conformal transformation, sometimes called a Weyl rescaling, thus moving to E frame in which the gravitational part is of the standard Einstein-Hilbert form.

None of realistic theories of gravity is conformally invariant.  Consequently physics looks different from frame to frame, though physical effects in different conformal frames can be related to each other unambiguously.  The latter fact is often expressed as ``equivalence" [\cite{d1}], though sometimes resulting in confusions.

A conformal transformation is a local change of units [\cite{d1}].  In
the context of Robertson-Walker cosmology, it is a time-dependent
change of the choice of the cosmic time, measured by different clocks.  In the prototype BD model with the scalar field decoupled from
matter {\em in the Lagrangian},
the time unit in J frame is provided by masses of matter particles, 
whereas the time in E frame is measured in units of the gravitational 
constant, or the Planck mass. As will be demonstrated  explicitly,
the way the universe evolves is quite different in the two frames.
We try to see how one can use this difference to select a particular frame.

We confine ourselves mainly to the analysis of the primordial nucleosynthesis which is
known to provide one of the strongest supports of the standard
cosmology.   We also focus on the simplest type of the theories which may apply only to the first step of the problem as stated above.  The result obtained here will still serve as a basis of more complicated models [\cite{hwnt}] to be applied to the second step.

Suppose first 
that at the onset of the whole process of nucleosynthesis the
universe had reached already the asymptotic phase during which it evolved according to the ``attractor'' solution for the BD model with $\Lambda$ added.  We find that, unlike in many analyses based on the
BD model {\em without} the
cosmological constant [\cite{wag}],  the physical result is acceptable in {\em neither} of the two frames for any
value of $\omega$, the well-known fundamental constant of the theory.
Conflicts with the standard picture are encountered also in the early
epoch just after inflation and in the dust-dominated era.

We then point out that the cosmological solution may likely show the
behavior of ``hesitation,'' in which the scalar field stays unchanged
for some duration.  If nucleosynthesis occurred during this phase,
both frames are equally acceptable, giving no distinction between
them.   Outside the era of nucleosynthesis, however, we inherit the
same conflicts for both conformal frames; hesitation itself offers no
ultimate solution.

Most of the conflicts can be avoided in E frame, as we find
fortunately, if, contrary to the original model, the 
scalar field is coupled to matter in a way which is not only simple 
and attractive from a theoretical point of 
view but is also roughly consistent with observations currently available.

In reaching this conclusion in favor of E frame, we emphasize that how
a time unit in certain conformal frame changes with time depends
crucially on how the scalar field enters the theory.  Searching for a
correct conformal frame is combined intricately with the search for a
theoretical model which would lead to a reasonable overall consistency with the cosmological observations.

In section 2, we start with defining the model first in J frame, then apply a
conformal frame moving to E frame.  We discuss in Section 3 the
attractor solution in some detail, including elaborated comparison
between the two conformal frames.  Section 4 discusses comparison with
phenomenological aspects of standard cosmology, particularly
primordial nucleosynthesis, dust-dominated universe and pre-asymptotic
era.  We enter discussion of the ``hesitation behavior" in Section 5,
still finding disagreement with standard cosmology.  To overcome the
difficulty we face, we propose in Section 6 a revision of the
theoretical model by abandoning one of the premises in the original BD
model, but appealing to a rather natural feature of scale invariance.  The
analysis is made both classically and quantum theoretically.  As we
find, the effect of quantum anomaly entails naturally the ``fifth force,'' featuring the finite-force range and violation of the weak equivalence principle (WEP).  Importance of further experimental studies is
 emphasized.     Final Section 7 is devoted to the concluding remarks.
Three Appendices are added for some details on (A) another attractor
solution, (B) mechanism of hesitation, and (C) loop integrals resulting in the anomaly.

\section{The model}

We start with the Lagrangian in J frame as given by
\beq
{\cal L}=\sqrt{-g}\left( \half\xi\phi^2 R -\half \epsilon g^{\mu\nu}
\partial_{\mu}\phi \partial_{\nu}\phi + \phi^{n}\Lambda +L_{\rm matter}
\right),
\label{cfr1-1}
\eeq
where our scalar field $\phi$ is related to BD's original notation $\varphi \equiv \phi_{BD}$ by
\beq
\varphi =\half\xi\phi^2,
\label{cfr1-2}
\eeq
also with the constant $\xi$ related to their $\omega$ by $\xi\omega =1/4$.  Notice that we use a unit system of $c = \hbar = 8\pi G (\equiv M^{-2}_{\rm P}) =1$.\footnote{In units of the Planck time ($2.71\times 10^{-43}$sec), the present age of the universe $\sim 1.4\times 10^{10}$y is given by $\sim 1.6\times 10^{60}$, while ``3 minutes" is $\sim 10^{45}$.}

We prefer $\phi$ defined above because by doing so we write equations in a form more familiar with conventional relativistic field theory, 
avoiding the {\em apparent} singularity $\varphi^{-1}$ in the kinetic term.  Also $\epsilon$ in \reflef{cfr1-1}) can be either of $\pm 1$ or 0.  Examples of $\epsilon =-1$ are provided by the dilaton field coming from 10-dimensional string theory [\cite{callan}], and the scalar field representing the size of internal space arising from compactifying $N$-dimensional spacetime with $N \geq 6$, while the latter with $N=5$ gives $\epsilon =0$.\footnote{A negative $\epsilon$ is found to be essential in the calculations in Refs. [\cite{dol}] carried out in J frame.}

Various models have been proposed for different choices of the
function of $\phi$ in the nonminimal coupling [\cite{wag}].  We adhere 
for the moment, however, to the simplest choice in \reflef{cfr1-1})
expecting it to be applied to situations of immediate physical
interests.  The factor $\phi^{n}$ is inserted because a cosmological constant in higher dimensions may appear in 4 dimensions multiplied with some power of $\phi$ depending on the model.

We assume RW metric, specializing to the radiation-dominated universe after the inflationary epoc, because it not only simplifies the calculation considerably but also applies to the era of nucleosynthesis.  Also choosing $k=0$ we obatin the equations
\beqa
6\varphi H^2&=& \epsilon\half \dot{\phi}^2 +\phi^{n}\Lambda +\rho_{r} -6 H\dot{\varphi},\label{cfr1-4}\\
\ddot{\varphi}+3H\dot{\varphi}&=&(4-n)\zeta^2 \phi^{n} \Lambda,\label{cfr1-5}\\
\dot{\rho}_{r} +4 H{\rho}_{r}&=&0,\label{cfr1-6}
\eeqa
where
\beq
H=\frac{\dot{a}}{a}.
\label{cfr1-7}
\eeq

Now consider a conformal transformation defined by
\beq
g_{*\mu\nu}=\Omega^2(x)g_{\mu\nu},\qquad{\rm or}\qquad ds^2_{*}=\Omega^2(x)ds^2.
\label{cfr1-8}
\eeq
By choosing
\beq
\Omega =\xi^{1/2}\phi,
\label{cfr1-9}
\eeq
we rewrite \reflef{cfr1-1}) as the Lagrangian in E frame:
\beq
{\cal L}=\sqrt{-g_{*}}\left( \half R_{*} - \half
	g^{\mu\nu}_{*}\partial_{\mu}\sigma\partial_{\nu}\sigma 
	+ V(\sigma)+L_{\rm *matter}  \right),
\label{cfr1-10}
\eeq
where
\beq
\phi=\xi^{-1/2}e^{\zeta\sigma},
\label{cfr1-11}
\eeq
with the coefficient $\zeta$ as defined by
\beq
\zeta^{-2} = 6+\epsilon\xi^{-1} = 2(3 +2\epsilon\omega),
\label{cfr1-12}
\eeq
and
\beq
V(\sigma)=\Omega^{-4}\phi^{n}\Lambda
=\Lambda\xi^{-n/2}e^{(n-4)\zeta\sigma}.
\label{cfr1-14}
\eeq
 Notice that the $\Lambda$ term in \reflef{cfr1-1}) now acts as a potential even with $n=0$.

We point out that the canonical field $\sigma$ is a normal field (not a ghost) if the right-hand side of \reflef{cfr1-12}) is positive:
\beq
6+\epsilon\xi^{-1} >0,
\label{cfr1-13}
\eeq
even with $\epsilon =-1$.  This positivity condition is satisfied
trivially for any finite value of $\xi$ if $\epsilon =0$ .  The
condition \reflef{cfr1-13}) should be obeyed even in the analysis in J
frame; in the presence of the nonminimal coupling that causes mixing
between $\phi$ and the spinless part of the tensor field, the sign of
the total scalar-tensor sector is not determined solely by $\epsilon$.
We notice that the conformal transformation serves also as the
relevant diagonalization procedure.  We impose $\xi > 0$, to keep the
energy of the tensor gravitational field positive.

With the RW metric also in E frame, the cosmological equations are
\beqa
&&3H^2_{*} =\rho_{*\sigma}+\rho_{*r},\label{cfr1-15} \\
&& \ddot{\sigma}+3H_{*}\dot{\sigma}+V'(\sigma)=0,\label{cfr1-16} \\
&&\dot{\rho}_{*r} +4 H_{*}{\rho}_{*r}=0,\label{cfr1-17}
\eeqa
where
\beq
\rho_{*\sigma}=\half \dot{\sigma}^2 +V(\sigma).
\label{cfr1-18}
\eeq
An overdot implies a differentiation with respect to $t_{*}$, the cosmic time in E frame defined by [\cite{fn1}]
\beqa
dt_{*}&=& \Omega dt,\label{cfr1-19}\\
a_{*}&=& \Omega a,\label{cfr1-20}
\eeqa
where the scale factor $a_{*}$ has also been introduced to define $H_{*}=\dot{a}_{*}/a_{*}$.


\section{Attractor solution}

The solutions of the cosmological equations can be obtained more easily in E frame than in J frame.  We in fact find a special solution;
\beqa
a_{*}(t_{*})&=&t^{1/2}_{*},  \label{cfr1-21}\\
\sigma(t_{*})&=&\bar{\sigma}+\frac{2}{4-n}\zeta^{-1}\ln t_{*},  \label{cfr1-22}\\
\rho_{*r}(t_{*})&=& \rho_{*r0} t^{-2}_{*},\quad\mbox{with}\quad\rho_{*r0}= \frac{3}{4}\left[ 1-\left( \frac{2}{4-n}\right)^2\zeta^{-2} \right], \label{cfr1-23}
\eeqa
where the constant $\bar{\sigma}$ is given by
\beq
\Lambda e^{(n-4)\zeta\bar{\sigma}}=\frac{1}{(n-4)^2} 
\xi^{n/2}\zeta^{-2}.
\label{cfr1-24}
\eeq
We also find 
\beq
\rho_{*\sigma}(t_{*})=\frac{3}{(n-4)^2}\zeta^{-2}t^{-2}_{*},
\label{cfr1-25}
\eeq
which is independent of $\Lambda$.

It is easy to see that, since $V(\sigma)$ is an exponentially
decreasing function if $n < 4$, $\sigma$ is pushed toward infinity,
hence giving $\rho_{*\sigma}\sim t_{*}^{-2}$ showing that $\Lambda_{\rm
eff}\equiv\rho_{*\sigma}$ does decay with time even with $n=0$, a purely
constant cosmological constant in 4 dimensions [\cite{fn1}].
According to \reflef{cfr1-22})  the
scalar field continues to grow; a unique feature that distinguishes our 
model from
other models [\cite{wag}] with the scalar
field designed to settle eventually to a constant.\footnote{If
$\Lambda$ is added to this type of models, this constant would yield a too large $\Lambda_{\rm eff}={\rm constant}$ which we attempt
to avoid without fine-tuning  parameters. }

To ensure a natural condition $\rho_{*r}>0$  we impose
\beq
\zeta^2 >\frac{4}{(4-n)^2}.\label{cfr1-25z}
\eeq
Though there is a subtlety as will be discussed in Appendix A, this is
one of the peculiar outcomes of including $\Lambda$.  According to
\reflef{cfr1-12}) this  predicts $\omega$ too small to be consistent
with the currently accepted lower bound.  Later, however, we will come
to propose a revised theoretical model in which $\zeta$ is not related
directly to phenomenological parameters.   We also notice from \reflef{cfr1-12}) that $\zeta >1/\sqrt{6}$ can be realized only for $\epsilon =-1$.\footnote{This is equivalent to choosing a negative $\omega$ according to some authors [\cite{kol2},\cite{santgreg}].}

\begin{figure}[t]
\hspace*{3.5cm}
\epsfxsize=7cm
\epsffile{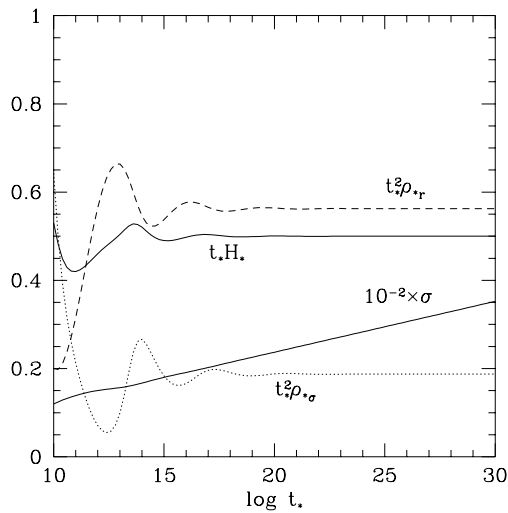}
\caption{An example  of the solutions to
\protect\reflef{cfr1-15})-\protect\reflef{cfr1-17})  with $n=0$.
We chose $\Lambda =\zeta =1.0$.  The initial values at $\log
t_{*1}=10$ are $\sigma =12.0, d\sigma /d\ln t_{*}=1.0, t_{*1}^2 \rho_{*r}=0.2$.  All the curves tend to the asymptotic lines
given by \protect\reflef{cfr1-21})-\protect\reflef{cfr1-23}) quickly after $\log t_{*} \sim 20$.  The present age of the universe is $\sim 10^{60}$.}
\end{figure}

It is already known in the literature that the solution \reflef{cfr1-21})-
\reflef{cfr1-25}) is an attractor, though many of the investigations
have been aimed at the question if it leads to a sufficient inflation
[\cite{halliw}-\cite{santgreg}].  In contrast we are interested in how
the  cosmological constant is relaxed in a manner consistent with
other known aspects of cosmology.\footnote{See also Ref. [\cite{yfq}]
for some of our earlier results.}  For this reason we start
integrating the equations sometime after the end of inflation when
sufficient amount of matter energy is created due to reheating, though
its details have not been well understood.  Fig. 1 shows how a typical
solution  of \reflef{cfr1-15})-\reflef{cfr1-17}) (with $n=0$) tends
asymptotically to the attractor behavior for $\log t_{*}\gsim 20$.  We emphasize that the behaviors are essentially the same for any values of $n$ as far as $n < 4$; according to \reflef{cfr1-14}) different $n$ is absorbed into different $\zeta$.

In this example we chose the initial value of $\sigma$ at the time
$t_{*1}(= 10^{10})$ in such a way that the resultant solution varies
``smoothly" around $t_{*1}$.  To be more specific, we require
$t_{*}H_{*}$, the local effective exponent in $a_{*}(t_{*})$, to remain of the order unity.   This is in accordance with assuming that the ``true" initial condition on the more fundamental level is given at a much earlier time.

The same attractor solutions can be obtained also in J frame.  This can be done most easily by substituting \reflef{cfr1-21})-\reflef{cfr1-25}) into \reflef{cfr1-19}) and \reflef{cfr1-20}).  After straightforward calculations we find
\beq
t= \left\{
\begin{array}{ll}
A_{n}t_{*}^{(2-n)/(4-n)}, &\qquad {\rm if}\hspace{2em} n\neq 2,\vspace{1em}\\
A_{2}\ln t_{*}, &\qquad {\rm if}\hspace{2em} n = 2,
\end{array}
\right.
\label{cfr1-26}
\eeq
and hence
\beq
a(t)= \left\{
\begin{array}{ll}
t^\alpha,\quad{\rm with}\quad \alpha =\mbox{{\Large$\half\frac{n}{n-2}$}}, &\qquad {\rm if}\hspace{2em} n\neq 2,\vspace{1em}\\
\exp \left( -\Lambda^2\xi^{-1/2}\zeta t \right), &\qquad {\rm if}\hspace{2em} n = 2,
\end{array}
\right.
\label{cfr1-27}
\eeq
where $A_{n}$ is a constant given in terms of $\Lambda, \xi$ and $n$.

Of special interest is the choice $n=0$, a purely constant
cosmological constant in 4 dimensions, giving
\beqa
t &=& A_{0}t_{*}^{1/2},\label{cfr1-28z}\\
a(t)&=& {\rm const}.
\label{cfr1-28}
\eeqa
\begin{figure}[t]
\hspace*{3.5cm}
\epsfxsize=7cm
\epsffile{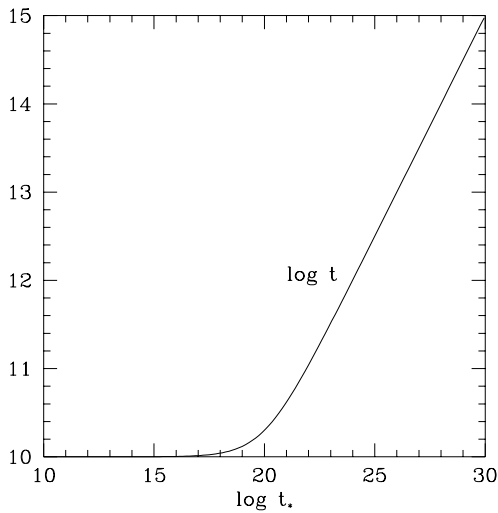}

\caption{The relation between two times, $t_{*}$ in E frame and $t$ in 
J frame, computed for the example shown in Fig. 1.}
\end{figure}
\begin{figure}[t]
\hspace*{3.5cm}
\epsfxsize=7cm
\epsffile{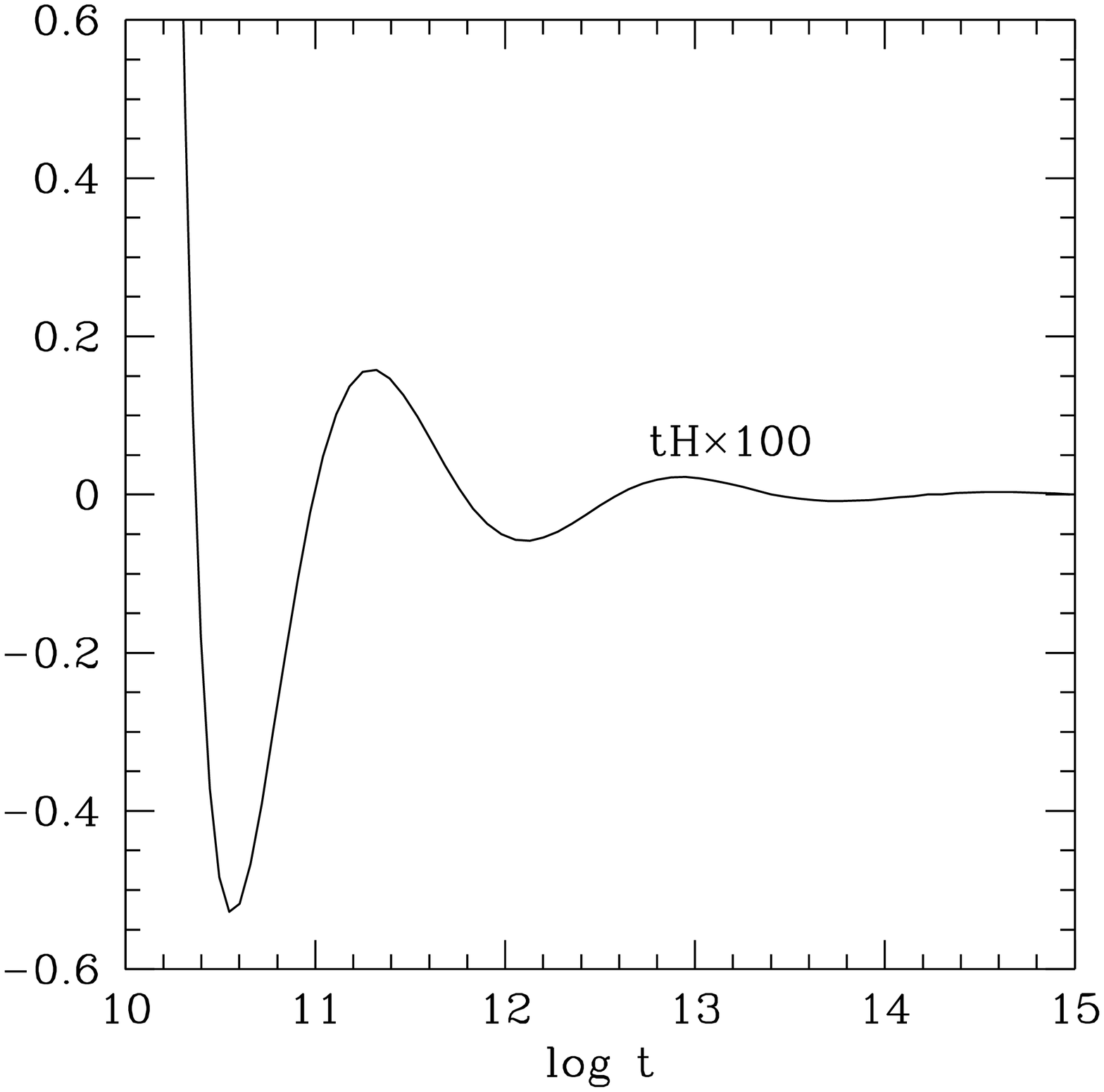}
\caption{$tH$ in J frame,  computed for the example shown in Fig. 1,
showing that the scale factor tends to a constant.  Exact behaviors are omitted for $t$ very close to the initial time.  See Fig. 4 for more details.}
\end{figure}

The asymptotic behavior \reflef{cfr1-28z}) is seen in Fig. 2 for the
same solution as in Fig. 1, under the condition $t_{1}=t_{*1}$ at the
``initial" time.   Notice that this behavior is preceded by a period
of an extremely slow take-off of $t(t_{*})$, which can be traced back
to a large value of $\Omega$ appearing in \reflef{cfr1-19}), related
to the remark stated above on the initial value $\sigma(t_{1*})$.  Fig. 3 shows how the asymptotic behavior
\reflef{cfr1-28}) is reached.  A closer look at the curve toward the
initial time reveals an oscillatory behavior too rapid to be drawn
here exactly.  This might have a disturbing effect, as will be discussed later.

Also for $n=2$, for which some of the exact solutions have been obtained [\cite{barrmaed},\cite{kol2},\cite{santgreg}], we find an exponentially contracting universe, though this can be converted to an exponentially expanding universe by reversing the direction of time.

The exponent $\alpha$ in the  first of \reflef{cfr1-27}) is larger than 1 for $ n>2 $, while it is negative for $0 <n <2$.  The exponent never reaches $1/2$ for any finite value of $n$; getting 0.45, for example, requires $n=-18$.  We also point out that the behavior differs considerably depending on whether the ``ordinary" matter ($\rho_{r}$ or $\rho_{*r}$) is included or not.

Summarizing, we expect substantial differences in the behavior of the
scale factor not only between the conformal frames but also among
different values of $n$ in J frame.  This demonstrates how crucial it
is to select a right conformal frame to discuss any of the physical
effects.  In the next section we analyze the physical implications to
be compared with the results of standard cosmology.


\section{Comparison with standard cosmology}

\subsection{Nucleosynthesis}

According to the standard scenario, light elements were created
through nuclear reactions in the radiation-dominated universe with the
temperature dropping proportional to the inverse square root of the
cosmic time.  The whole process is analyzed in terms of
nonrelativistic quantum mechanics in which particle masses are taken
obviously as {\em constant}.  In scalar-tensor theories, on the other hand,
particle masses depend generically on the scalar field, hence on time.
The prototype BD theory is unique in that masses are true
constants thanks to the assumption that the scalar field is decoupled
from matter in the Lagrangian.\footnote{This is a major point that
distinguishes between BD model and the one due to Jordan [\cite{jd}] who
pioneered the nonminimal coupling.}  For this reason the mass $m$ of a particle is a pure constant in J frame, specifically denoted by $m_{0}$.

We should recall that we have so far considered relativistic matter alone.  One may nevertheless include massive particles which play no role to determine the overall cosmological evolution, but may serve to provide standards of time and length.

To find how masses become time-dependent after a conformal transformation, it is sufficient to consider a toy model of a free massive real scalar field $\Phi$, described by the matter Lagrangian in J frame;
\beq
{\cal L}_{\rm matter}=\sqrt{-g}\left(  -\half g^{\mu\nu}\partial_{\mu}\Phi\partial_{\nu}\Phi -\half m_{0}^2 \Phi^2 \right),
\label{cfr1-29}
\eeq
where no coupling to $\phi$ is introduced.  After the conformal transformation \reflef{cfr1-8}), we find
\beq
{\cal L}_{\rm matter}=\sqrt{-g_{*}}\left(  -\half \Omega^{-2}g^{\mu\nu}_{*}\partial_{\mu}\Phi\partial_{\nu}\Phi -\half \Omega^{-4}m_{0}^2 \Phi^2 \right).
\label{cfr1-29a}
\eeq

The kinetic term can be made canonical in terms of a new field $\Phi_{*}$ defined by
\beq
\Phi =\Omega \Phi_{*},
\label{cfr1-30}
\eeq
thus putting \reflef{cfr1-29a}) into 
\beq
{\cal L}_{\rm matter}=\sqrt{-g_{*}}\left(  -\half g^{\mu\nu}_{*}{\cal D}_{\mu}\Phi_{*}{\cal D}_{\nu}\Phi_{*} -\half m_{*}^2 \Phi^2_{*} \right),
\label{cfr1-30a}
\eeq
where
\beqa
{\cal D}_{\mu}\Phi_{*}&=& \left[ \partial_{\mu} +\zeta \left( \partial_{\mu}\sigma \right)  \right]\Phi_{*},\label{cfr1-30b}\\
\mbox{}\nnb\\
m_{*}^2&=& \Omega^{-2}m_{0}^2.
\label{cfr1-30c}
\eeqa
We point out that this relation holds true generally beyond the simplified model considered above.

From \reflef{cfr1-9}), \reflef{cfr1-11}) and \reflef{cfr1-22}) we find
\beq
\Omega \sim \phi \sim t_{*}^{2/(4-n)},
\label{cfr1-33}
\eeq
hence giving a {\em time-dependent mass} in E frame.  This would
result in the reduction of masses as much as
$1-1/\sqrt{10}\approx$70\% if $n=0$, for example, in the period
$100\,$-$1000\,$sec, during which major part of synthesis of light
elements is supposed to have taken place with the temperature dropping
in the same rate.  Obviously this is totally in conflict with the success of the standard scenario.

In this way we come to a dilemma; J frame is selected uniquely because of constancy of masses as taken for granted in conventional quantum mechanics to analyze the physical processes,
while E frame is definitely preferred to have the universe that cooled down sufficiently for light elements to form.

In passing we offer a simple intuitive interpretation on the relation
between the two conformal frames.  In the present context, the time unit
$\tau$ in E frame is provided by $m_{*}^{-1}\sim \Omega$.  The time
$\tilde{t}$ measured in units of $\tau$ may be defined by $d\tilde{t}
= dt_{*}/\tau \sim \Omega^{-1}dt_{*}$.  Comparing this with
\reflef{cfr1-19}), we find $\tilde{t}=t$.  On the other hand, the only
dimensionful constant in E frame is $M_{\rm P}$.  In this sense E
frame corresponds to the time unit provided by the gravitational
constant.  For $n=0$, for example, the microscopic length scale
provided by $m_{*}^{-1}$ expands in the same rate as the scale factor
$a_{*}(t_{*})$, hence showing no expansion in $a(t)$ in \reflef{cfr1-20}).

\subsection{Dust-dominated era}

In the standard theory the radiation-dominated era is followed by the
dust-dominated universe.  Its description is, however, likely
problematic, as will be shown.

Due to the equation
\beq
\BBox_{*}\sigma = c\, \zeta T_{*},
\label{cfr1-81}
\eeq
with $T_{*}=-\rho_{*d}$, the right-hand side of \reflef{cfr1-16}) acquires an additional term
\beq
c\,\zeta\rho_{*d},
\label{cfr1-82}
\eeq
where, for the non-relativistic matter density $\rho_{*d}$, $c=1$ for the prototype BD model, but we allow $c$ to be different in the proposed revision which will be discussed later.  Suppose the total matter density is the sum $\rho_{*r}+\rho_{*d}$, which would replace $\rho_{*r}$ in \reflef{cfr1-15}).  Corresponding to \reflef{cfr1-17}), we find\beq
\dot{\rho}_{*d} +3 H_{*}{\rho}_{*d}= -c\,\zeta\rho_{*d}\dot{\sigma},
\label{cfr1-83}
\eeq
where the right-hand side is included to meet the condition from the Bianchi identity.

We obtain the attractor solution with
\beq
a_{*}(t_{*})=t_{*}^{\alpha_{*}}, \quad \mbox{with}\quad
\alpha_{*}=\frac{4-c}{6},
\label{cfr1-84}
\eeq
and 
\beq
\rho_{*d0}=\frac{4-c}{3}-\frac{1}{4}\zeta^{-2}.
\label{cfr1-85}
\eeq
Equation \reflef{cfr1-22}) still holds true.  Notice, however, that  \reflef{cfr1-84}) gives $\alpha_{*}=1/2$ for $c=1$, the same behavior as in the radiation-dominated era.  This seems ``uncomfortable," if we wish to stay close to the realm of the standard scenario, though we may not entirely rule out a highly contrived way for a reconciliation.  On the other hand, we would obtain the conventional result $\alpha_{*}=2/3$ for $c=0$.

We also find that the relation $t\sim t_{*}^{1/2}$ remains unchanged
(for $n=0$), and then follows $a =\mbox{const}$ in J frame again for
$c=1$.  This is certainly disfavored, as in the analysis of
nucleosynthesis.

\subsection{Pre-asymptotic era}

As we noticed in Fig. 3, $H$ in J frame seems to be oscillatory around zero in early epochs, arousing suspicion of a contracting universe.    More details towrad $t_{1}=t_{*1}$ are  shown in
Fig. 4, in which $tH$ is plotted against $\log t_{*}$ instead of 
$\log t$; the dip of
$tH$ would be too sharp to be shown if plotted against $\log t$.  We
also plotted $\log a$, which does show a decrease of the scale factor
in J frame.  The example here shows that the scale factor $a$ which
comes to at rest toward the asymptotic era is even smaller than at $t_{*1}$.
The exact amount of contraction depends on the choice of the
parameters, still making it considerably difficult to reach a compromise with the idea that the early universe had cooled down sufficiently to trigger the process of nucleosynthesis. 
\begin{figure}[tbh]
\hspace*{3.5cm}
\epsfxsize=7cm
\epsffile{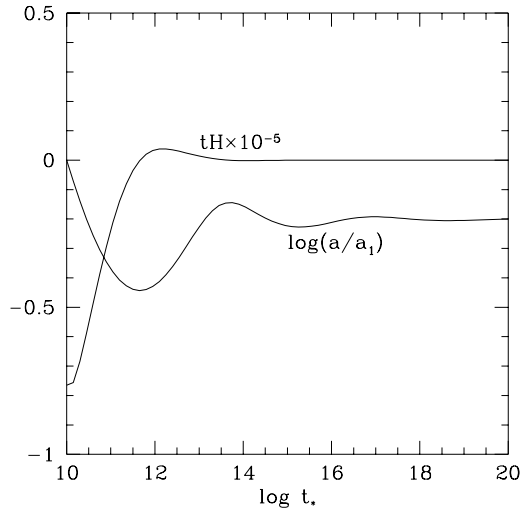}

\caption{More detailed behavior of $tH$ and $\log a$ plotted against
$\log t_{*}$, computed for the example shown in Fig. 1.   The value of the scale factor $a$ in J frame when it
comes to at rest toward the asymptotic behavior is even
{\em smaller} than $a_{1}$ at the ``initial'' time.  }
\end{figure}


\section{Hesitation behavior}

We have so far concentrated on the attractor solution, to which some
solutions, depending on the initial conditions, do tend smoothly, as
demonstrated in Fig. 1, for example.  We point out, however,
there is an important pattern of deviation from this smooth behavior.
As illustrated in Fig. 5, the scalar field may remain almost at rest temporarily before entering the
asymptotic phase in which it resumes to increase to approach the
attractor solution.  This ``hesitation" behavior may occur if
$\rho_{*r}\ll \rho_{*\sigma}$ at the initial time, as will be shown in
detail in 
Appendix B [\cite{fon}].  

\begin{figure}[tbh]
\hspace*{3.5cm}
\epsfxsize=7cm
\epsffile{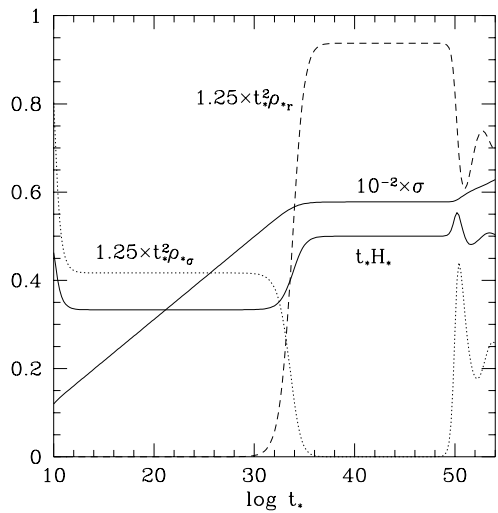}
\caption{An example of a hesitation behavior, obtained by choosing
the initial value of $t_{*}^2 \rho_{*r}$ as $2.0 \times 10^{-15}$, with other
parameters the same as in the solution in Fig. 1.  
The $K_{\sigma}$-domination occurs for $11<\mbox{\hspace{-1.2em}\protect\raisebox{-.9ex}{$\sim$}}\log t_{*}<\mbox{\hspace{-1.2em}\protect\raisebox{-.9ex}{$\sim$}}\:30$, while the
hesitation behavior is seen clearly for $35 <\mbox{\hspace{-1.2em}\protect\raisebox{-.9ex}{$\sim$}}  \log t_{*} <\mbox{\hspace{-1.2em}\protect\raisebox{-.9ex}{$\sim$}}49$, 
followed by the usual radiation-dominated universe. Notice that ``3 minutes" corresponds to $\sim 10^{45}$.}
\end{figure}
\begin{figure}[tbh]
\hspace*{3.5cm}
\epsfxsize=7cm
\epsffile{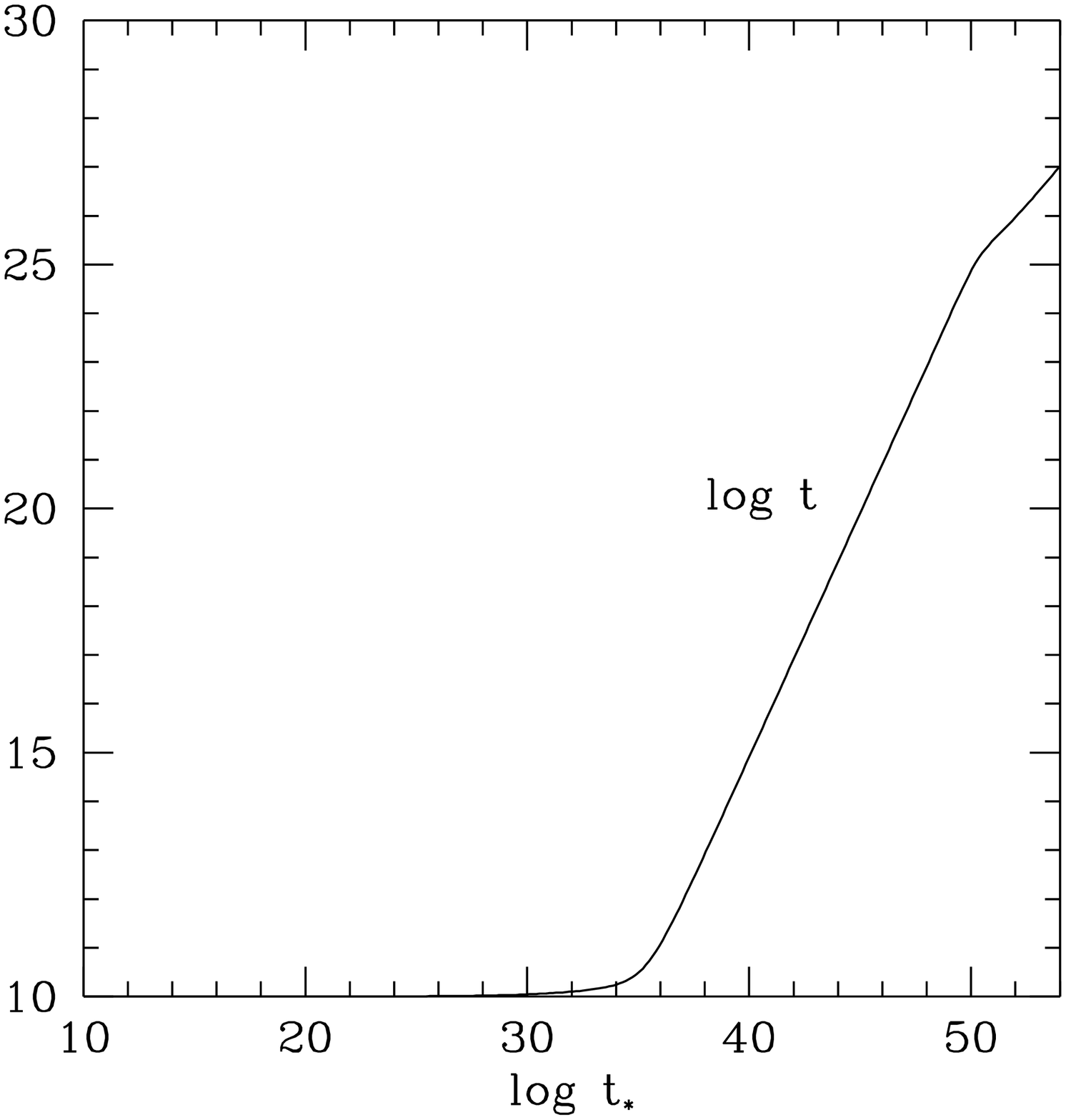}
\caption{The relation between the two times for the solution shown in
Fig. 5.  The behaviors $t\sim t_{*}$  and
$t\sim t_{*}^{1/2}$ are seen for $35 <\mbox{\hspace{-1.2em}\protect\raisebox{-.9ex}{$\sim$}}  \log t_{*} <\mbox{\hspace{-1.2em}\protect\raisebox{-.9ex}{$\sim$}}49$ and $\log
t_{*}> \mbox{\hspace{-1.2em}\protect\raisebox{-.9ex}{$\sim$}}50$, respectively.}

\end{figure}
\begin{figure}[tbh]
\hspace*{3.5cm}
\epsfxsize=7cm
\epsffile{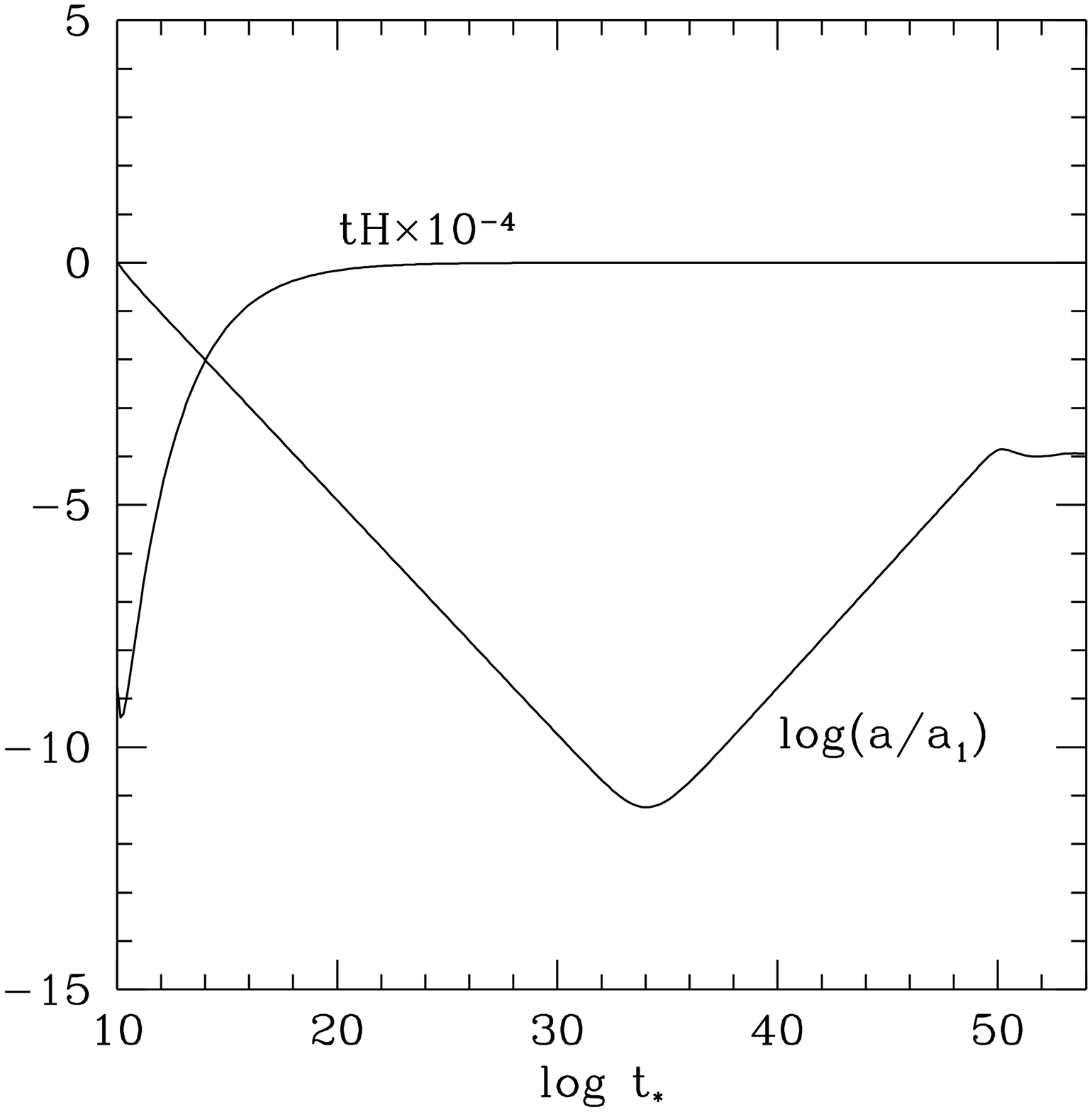}
\caption{Like in Fig. 4 for the solution without hesitation, the solution 
shown in Fig. 5 results in the considerable amount of contraction of
the scale factor in J frame in the early universe. }
\end{figure}

Notice that $t_{*}H_{*}$, which equals the effective exponent
$\alpha_{*}$ if the scale factor is approximated locally by
$a_{*}(t_{*})\sim t_{*}^{\alpha_{*}}$, tends to 1/2 after some
wiggle-like behavior that separates the plateau of the same value 1/2
in the hesitation period.  See also Appendix B for the mechanism behind another plateau of $t_{*}H_{*}=1/3$.

With the scalar field  nearly constant during this hesitation phase, particle
masses are also nearly constant, and all the other cosmological 
effects are virtually the same as those in the standard theory as
long as $\rho_{*\sigma}\ll\rho_{*r}$, as is the case for $\log t_{*}\gsim 35$ in the example of Fig. 5, in 
which we chose the parameters in such a way that the hesitation period
$\log t_{*}= 35\,\mbox{-}48$ covers the era of nucleosynthesis.  Moreover,
the constant scalar field makes the conformal 
transformation  \reflef{cfr1-19}) and \reflef{cfr1-20}) trivial,
implying that the two conformal frames are essentially equivalent to
each other.

In more details, however, we find some differences between them.  In our example, we carried out the transformation
\reflef{cfr1-19}) and \reflef{cfr1-20}), showing first in Fig. 6, how the two time variables $t_{*}$ and $t$ are related to each other; the period of $t\sim t_{*}$ during hesitation is present in addition to the behavior as shown in Fig. 2 without the hesitation behavior.

As in Fig. 4, we find in Fig. 7 that the universe in J frame had experienced a
considerable contraction prior to the epoch of  nucleosynthesis, making the scenario of the evolution in early epochs desperately different from the standard theory.  See Appendix B for its origin.

 No such problems will occur if we are still during the hesitation period at the present time.  If this happens, however, no distinction is present between the two frames, offering no obstruction to choose E frame either.  From this point of view, we do not consider this possibility any further.

One might suspect that all of these ``conflicts" with the standard picutre come directly from the theoretical models to start with.  In fact it seems obvious that we would be in a much better position if particle masses were constant in E frame rather than in J frame.  This can be achieved, as we will show, by modifying one of the assumptions in the prototype model in a rather natural manner.

\section{A proposed revision}

\subsection{Classical thoery}
Unlike in the original BD model, let us start with the matter Lagrangian in J frame:
\beq
{\cal L}_{\rm matter}=\sqrt{-g}\left(  -\half g^{\mu\nu}\partial_{\mu}\Phi\partial_{\nu}\Phi - \half f^2 \phi^2 \Phi^2 
-\frac{1}{4!}\lambda\Phi^4 \right),
\label{cfr1-33b}
\eeq
in place of \reflef{cfr1-29}).  We introduced the coupling of $\phi$
to the matter field $\Phi$ [\cite{fn1},\cite{yf3}], hence abandoning one of the premises in
the original BD theory.   Also there is no mass term of $\Phi$;
``mass" of the field $\Phi$ is $f\phi$ which is no longer constant.
With the choice \reflef{cfr1-33b}), therefore, J frame loses its
privilege to be a basis of the theoretical analysis of
nucleosynthesis.  We also introduced the self-coupling of $\Phi$ to
illustrate the effect of the quantum anomaly.

After the conformal transformation we obtain the same Lagrangian
\reflef{cfr1-30a}) (plus the self-coupling term) but with
\reflef{cfr1-30c}) replaced by 
\beq
m_{*0}^2= f^2\phi^2 \Omega^{-2} = \xi^{-1}f^2,
\label{cfr1-34}
\eeq
which is obviously constant, making E frame now a relevant frame for realistic cosmology.  We point out that no matter coupling of $\sigma$ is present in E frame.  Absence of the coupling in the Lagrangian having no nonminimal coupling implies a complete decoupling, unlike the corresponding situation in J frame in the prototype model.  This corresponds to the choice $c=0$ in \reflef{cfr1-81})$\,$- \reflef{cfr1-85}), thus leaving the results in E frame the same as those of standard cosmology also in dust-dominated universe.  On the other hand, the scalar field can be detected in no ways by measuring its contribution to the force between matter objects, or the conventional tests of General Relativity, hence removing the constraints on $\omega$ (or $\xi$) obtained so far.  Its effect may still be manifest through cosmological phenomena.

The time scale in E frame is provided {\em commonly} by particle masses and the gravitational coupling constant.  This implies that no time variability of the gravitational constant should be observed if measured by atomic clocks with their unit given basically by particle masses.\footnote{Strictly speaking, the time unit of atomic clocks depends also on the fine-structure constant, whose constantcy is assumed, however, at this moment.}

The above scheme is attractive because the coupling constant $f$ is {\em dimensionless}, hence vesting scale invariance in the gravity-matter system except for the $\Lambda$ term.  By applying Noether's procedure we obtain the dilatation current as given by
\beq
J^{\mu}=\half\sqrt{-g_{*}}g^{\mu\nu}_{*}\left[ 2\zeta^{-1}\partial_{\nu}\sigma +\left( \partial_{\nu}+2\zeta \partial_{\nu}\sigma \right)\Phi_{*}^2 \right],
\label{cfr1-73b}
\eeq
which is shown to be conserved by using the field equations.  This
conservation law remains true even after the conformal transformation,
with a nonzero mass as given by \reflef{cfr1-34}).  This implies that
the scale invariance is broken {\em spontaneously} due to the trick by which
a dimensionful constant $M_{\rm P}(=1)$ has been ``smuggled" in
\reflef{cfr1-9}).  In this context $\sigma$ is a Nambu-Goldstone
boson, a dilaton.  This invariance together with constancy of particle
masses will be lost, however, if one includes quantum effects due to
the non-gravitational couplings {\em among} matter fields, as will be sketched below.

\begin{figure}[tbh]
\hspace*{3.5cm}
\epsfxsize=7cm
\epsffile{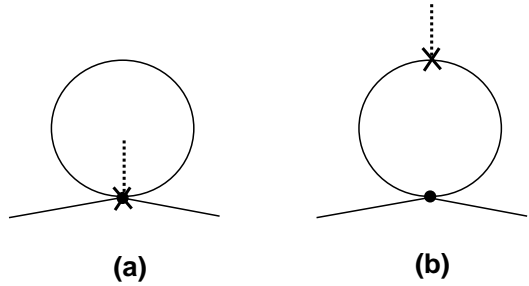}
\caption{One-loop diagrams giving an anomaly.  The blobs are for the
coupling $(\lambda /4!)\Phi^4$, while the crosses represent a coupling to 
 $\sigma$, the linear term in \protect\reflef{cfr1-33b2}).   }
\end{figure}

\subsection{Quantum anomaly}
Consider one-loop diagrams for the coupling between $\Phi^2_{*}$ and $\sigma$ assumed to carry no momentum, as illustrated in Fig. 8, arising from the non-gravitationa coupling  $(\lambda /4!)\Phi^4$.  They will result, according to quantum field theory, generally in divergent integrals, which may be regularized by means of continuous spacetime dimensions.  Corresponding to this, we rewrite previous results extended to $N$ dimensions. Equation \reflef{cfr1-30}) is then modified to 
\beq
\Phi =\Omega^{\nu -1} \Phi_{*},
\label{cfr1-35}
\eeq
where $\nu =N/2$.  Also the relation \reflef{cfr1-9}) is changed to 
\beq
\Omega = \xi^{1/(N-2)}\phi^{1/(\nu -1)}.
\label{cfr1-36}
\eeq
The Lagrangian \reflef{cfr1-33b}) is now put into
\beq
{\cal L}_{\rm matter}=\sqrt{-g_{*}}\left(  -\half g^{\mu\nu}_{*}{\cal D}_{\mu}\Phi_{*}{\cal D}_{\nu}\Phi_{*} - S  \right),
\label{cfr1-33b1}
\eeq
where ${\cal D}_{\mu}$ is given by \reflef{cfr1-30b}) while $S$ is defined by
\beq
S=e^{2(\nu -2)\zeta\sigma}S_{0},
\label{cfr1-33b2}
\eeq
with
\beq
S_{0}=\xi^{-1}\frac{f^2}{2}\Phi_{*}^2+\frac{1}{4!}\lambda\Phi^4_{*}. 
\label{cfr1-33b3}
\eeq

As a result the right-hand side of \reflef{cfr1-34}) is multiplied by $e^{(N-4)\zeta\sigma}$, which might be expanded with respect to $(N - 4)\zeta\sigma$ according to \reflef{cfr1-11}); decoupling occurs {\em only in 4 dimensions}.  Following the rule of dimensional regularization, we keep $N$ off 4 until the end of the calculation including the evaluation of loop integrals which are finite for $N\neq 4$.

The divergences coming from the $\Phi_{*}$ loops are
represented by poles $(N-4)^{-1}$ which cancel the factor $N-4$ that multiplies the $\sigma$ coupling as stated above, hence yielding an effective interaction of the form
\beq
-L_{\sigma\Phi}' =  g_{\sigma}\frac{m^2_{*0}}{M_{P}}\Phi_{*}^2\sigma,
\label{cfr1-36a}
\eeq
for $N\rightarrow 4$, where the coupling constant $g_{\sigma}$ is given by
\beq
g_{\sigma}=\zeta\frac{\lambda}{8\pi^2},
\label{cfr1-36b}
\eeq
with the details of computation presented in Appendix C.  In \reflef{cfr1-36a}) we reinstalled $M_{P}^{-1}=\sqrt{8\pi G}$ to remind that this coupling is basically as weak as the gravitational interaction.

Suppose $\Phi_{*}$ field at rest is a representative of dust matter.  We may then take $\rho_{*d}\approx m_{*0}^2\Phi_{*}^2$.  Comparing \reflef{cfr1-36a}) with \reflef{cfr1-81}) we find
\beq
g_{\sigma}=-c\zeta.
\label{cfr1-36bb}
\eeq
In this way we obtain  $g_{\sigma}$, and hence $c$, which are {\em nonzero finite} due to a non-gravitational interaction.

It should be warned that our determination of $c$ in \reflef{cfr1-81})
never implies that $\sigma$ couples to the trace of the
energy-momentum tensor.  From \reflef{cfr1-36b}) we find that the
coefficient $g_{\sigma}$ depends on $\lambda$ which is {\em not
related} to the mass directly, hence may differ from particle to
particle.  If we include many matter particles, the right-hand side of
\reflef{cfr1-81}) should be the sum of corresponding components with
different coefficients.  For this reason the force mediated by
$\sigma$ {\em fails generically to respect WEP}.  This was shown more
explicitly in our QED version [\cite{yf3}].

We notice that the above calculation is essentially the same as those
by which various ``anomalies" are derived, particularly the trace
anomaly [\cite{ce}].  The relevance to the latter can be shown explicitly if  we consider the source of $\sigma$  in the limit of weak gravity.

We first derive
\beq
\partial_{\mu}J^{\mu}=2\zeta (\nu -2)\sqrt{-g_{*}}S,
\label{cfr_88a}
\eeq 
indicating that scale invariance is broken {\em explicitly} for $N\neq 4$.  Now deriving \reflef{cfr1-36a}) is essentially calculating the quantum theoretical expectation value of $S_{0}$ between two 1-particle states of $\Phi_{*}$;
\beq
< 2\zeta (\nu -2)S_{0} >_{\Phi_{*}} = g_{\sigma} m_{*0}^2,
\label{cfr_88b}
\eeq
ignoring terms higher order in $\zeta (=\zeta M^{-1}_{\rm P})$.  Combining this with \reflef{cfr1-81}) we may write
\beq
\BBox_{*}\sigma = <\partial_{\mu}J^{\mu}>_{\Phi_{*}}.
\label{cfr1-73a}
\eeq
This equation shows that $\sigma$ couples to breaking of scale
invariance effected by quantum anomaly,\footnote{In this respect we rediscover the
proposal by Peccei, Sola and Wetterich [\cite{wett}], though their
approach to the cosmological constant problem is different from ours.} though this simple realtion is justified only up to the lowest order in $M^{-1}_{\rm P}$.

One might be tempted to extend the analysis by identifying $\Phi$ with the Higgs boson in the standard electro-weak theory or Grand Unified Theories, hence predicting observable consequences.  As we find, however, the realistic analysis is ought to be more
complicated; we must take contributions from other couplings including the Yukawa and QCD interactions into account. Even more serious is
that we are still short of a complete understanding of the ``content'' of
nucleons, the dominant constituent of the real world.  We neverthelss
attempt an  analysis, as will be sketched briefly, leaving further details to the future publications.

From a practical point of view, we need the coupling strength of
$\sigma$ to nucleons, through the
couplings to quarks and gluons.  By a calculation parallel to that
leading to \reflef{cfr1-36b}) we obtain\footnote{The coefficient 6 in eq. (52) of Ref. [\cite{yf3}] should be replaced by 15/4. (Also $\beta$ is our present $\zeta$.)  The QCD result, our \protect\reflef{cfr1-73c}), is obtained by multiplying further by the factor $(N^2 -1)/2N$ which is 4/3 for $N=3$.}
\beq
c_{q}=-\frac{5\alpha_{s}}{\pi}\sim 0.3,
\label{cfr1-73c}
\eeq 
for a $\sigma$-quark coupling, where $\alpha_{s}$ is the QCD analog of
the fine-structure constant, most likely of the order of $\sim 0.2$.  We then evaluate 
\beq
c_{N}m_{N}= c_{q}<\sum_{i}m_{i}\bar{q}_{*i}q_{*i} >_{N},
\label{cfr1-73c1}
\eeq
where the nucleon matrix element has been estimated to be $\sim 60$MeV, [\cite{loch}] which together with \reflef{cfr1-73c}) would give
\beq
c_{N}\sim 1.8\times 10^{-2}.
\label{cfr1-73c2}
\eeq
It is interesting to note that this is rather close to  the constraints obtained from observations, as will be shwon.

\subsection{Fifth force and cosmology}
By replacing  $\zeta$ in \reflef{cfr1-12}) by $c\zeta$ also choosing
$\epsilon =+1$  in accordance with the conventional analysis, we obtain
\beq
c^{-2}\zeta^{-2}=8(3+2\omega),
\label{cfr1-71}
\eeq
or
\beq
c\zeta \sim\frac{1}{4\sqrt{\omega}} \lsim 0.8\times 10^{-2},
\label{cfr1-72}
\eeq
if we accept $\omega \gsim 10^3$ obtained from the solar-system experiments, assuming the force-range of $\sigma$  longer than 1 solar unit.  Notice that $\sigma$ is now a pseudo Nambu-Goldstone boson which likely acquires a nonzero mass.  Combining this with the constraint \reflef{cfr1-25z}), we find
\beq
|c| \lsim 1.6\times 10^{-2},
\label{cfr1-72a}
\eeq
which we find is  nearly the same as \reflef{cfr1-73c2}).

Similar constraints may come from the ``fifth force" phenomena which are characterized by a finite force-range and WEP violation both of which are generic in the present model.  The parameter $\alpha_{5}$, the relative strength of the fifth force, is given by
\beq
\alpha_{5}\sim 2g_{\sigma}^2 \sim c_{N}^2,
\label{cfr1-74}
\eeq
expecting that the coupling comes mainly from the one to nucleons.

Since $g_{\sigma}$ may depend on the object $\sigma$ couples to, as was pointed out, $\alpha_{5}$ may also depend on the species of nuclei, for example, between which $\sigma$ is exchanged, to be denoted by $\alpha_{5ij}$.

From the observations carried out so far, we have the upper bounds given roughly by [\cite{fsb}]
\beq
|\alpha_{5ij}|\lsim 10^{-5}.
\label{cfr1-75}
\eeq
In view of the fact that the result depends crucially on the assumed value of the
force-range as well as the model of WEP violation, we may consider that the estimate \reflef{cfr1-74}) with \reflef{cfr1-73c2}) is approximately consistent with \reflef{cfr1-75}), hence providing a renewed motivation for further studies of the subject both from theoretical and experimental sides.

The analysis is still tentative particularly because \reflef{cfr1-73c}) is justified only to the lowest order with respect to $\alpha_{s}$, though the renormalization-group technique can be used to include the leading-order terms.  Potentially more important would be to estimate the contribution from gluons.  Corresponding to the second term of \reflef{cfrc_1}), we should include the direct coupling of $\sigma$ to the QCD coupling constant $g_{s}$ as given by
\beq
 \lim_{\nu\rightarrow 2}\zeta (\nu -2)g_{s}Z_{g}\sigma =2\pi\zeta\beta_{0}g_{s}\alpha_{s}\sigma,
\label{cfr1-75a}
\eeq
where $\beta_{0}=(4\pi)^{-2}(11 -2n_{f}/3)$ with $n_{f}$ the number of flavors.  It is yet to be studied how this coupling would affect the simple result \reflef{cfr1-73c2}) through the gluon content of a
nucleon.

It should also be emphasized that the right-hand side of
\reflef{cfr1-73c1}) (even with the modification stated above) is quite different from the matrix element of the conventional energy-momentum tensor or its trace related directly to observations; only the anomalous part participates.\footnote{An analysis due to Ji [\cite{xj}] showing that the trace anomaly contributes approximately 20\% of the nucleon mass might also be suggestive.}

One might argue that $c_{N}$ which would be too large to be allowed by
the phenomenological constraints could emerge if we apply the same
type of calculation to a nucleon considered to be an elementary
particle, as was attempted in the simpler QED version [\cite{yf3}].
We point out, however, the finite size of a composite nucleon would serve to suppress ultra-violet divergences, thus failing to produce an anomaly, which is a manifestation that the underlying theory is divergent.

It is rather likely that $\sigma$ couples also to the nuclear binding energy which is generated supposedly by the exchanged mesons.  This would make the analysis of composition-dependence even more complicated .\footnote{Most of the past analyses on the composition-dependent experiments [\cite{adel}] have been made based on the assumption that the fifth force is decoupled from the nuclear binding energy, which might be too simplified from our point of view.}

 We now turn to cosmological aspects.  Adding \reflef{cfr1-36a}) to the ``classical" mass term, we typically obtain
\beq
-L'_{m\Phi}=\half m_{*}^2 \Phi_{*}^2,\quad \mbox{with}\quad 
m_{*}^2=  m_{*0}^2( 1-c\zeta\sigma +\cdots ).
\label{cfr1-71z}
\eeq
We focus on the cosmological background $\sigma(t_{*})$ rather than the sapce-time fluctuating part which would mediate a force between matter objects as considered above.  In this sense $m_{*}$ depends on time.  We must then apply another conformal transformation to cancel this effect.  If, however, $c$ is sufficiently small, as indicated in \reflef{cfr1-72a}), one may expect that the expansion in \reflef{cfr1-71z}) is exponentiated giving
\beq
m_{*}(\sigma) \approx m_{*0} e^{-c\zeta\sigma}\sim t^{-c/2}_{*},
\label{cfr1-71y}
\eeq
where we have used the asymptotic behavior \reflef{cfr1-22}) with
$n=0$.  We expect \reflef{cfr1-71y}) holds true approximately for realistic nucleons or nuclei.

Notice that the final expression on the time variation is independent of $\zeta$.  The
resulting conformal frame is expected to be close to E frame.  A small $|c|$ is also favored from \reflef{cfr1-84}) for the attractor
solution in the dust-dominated unvierse.  It would further follow that $\dot{G}/G$ is somewhat below the level of $10^{-10}{\rm y}^{-1}$, in accordance with the observations [\cite{gdot}].

If, on the contrary, $|c|$ is ``large," we may even not be able to
compute the required conformal transformation unless we determine
higher-order terms in the parenthesis in \reflef{cfr1-71z}).  All in
all, we would be certainly ``comfortable" if $|c|$ is sufficiently
small.  On the other hand, it seems unlikely that $|c|$ is smaller
than unity by many orders of magnitude because we know no basic reason
why it should be so.  The present constraint \reflef{cfr1-72a}) might
be already close to the limit which one can tolerate in any reasonable theoretical calculation.  In this sense, probing $\alpha_{5}$ with accuracy improved by a few orders of magnitude would be crucailly important to test the proposed model of broken scale invariance.  If we come to discover any effect of this kind, it would provide us with valuable clues on how nucleons and nuclei are composed of quarks and gluons.

\section{Concluding remarks}

Having introduced a scalar field in order to relax the cosmological
constant within the realm of the standard scenario, we come to a conclusion:  At the classical level, E frame is the only
choice provided the J frame version has a scale invariant coupling
between the scalar field and the matter fields without the intrinsic
mass terms.  Most crucial are constancy of particle masses and the
expansion law of the universe during the epoch of primordial
nucleosynthesis.  A quantum anomaly serves naturally to break scale
invariance explicitly, offering yet another support for the occurrence of the fifth force featuring WEP violation.  A tentative calculation based on QCD yields the coupling strength roughly consistent with the observational upper bounds.  The physical conformal frame should then remain
close to E frame.  Improved efforts to probe the fifth force is
encouraged, though detailed theoretical predictions are yet to be
attempted.

We point out, however, there is a possible way to leave $\sigma$ completely decoupled even with quantum effects included.  We may demand that the $\phi$-matter coupling in \reflef{cfr1-33b}) has a coupling constant $f$ which is dimensionless {\em in any dimensions}.  This can be met if we replace the second term in the parenthesis in \reflef{cfr1-33b}) by
\beq
-\half f^2 \phi^{2/(\nu -1)}\Phi^2,
\label{cfr1-41}
\eeq
resulting in $m_{*}$ which is shown to be completely
$\sigma$-independent in E frame, even with the quantum effect
included.

We find, on the other hand, that the term of the nonminimal coupling
of the form  $\phi^2 R$, as in the first term of \reflef{cfr1-1}), is
multiplied {\em always} with a dimensionless constant for any
dimensions. This is a fact that underlies the whole discussion of
scale invariance, making another difference from the models
[\cite{wag}] which allow more general functions of the scalar field.
Scale invariance is respected also in a new approach to the
scalar-tensor theory based on $M_{4}\times Z_{2}$ [\cite{saito}].

The present study is limited because the model with a single scalar
field might be too simple to account for a possible nonzero
cosmological constant.\footnote{See, for example, [\cite{hwnt}] for a model with another
scalar field implementing a ``sporadic'' occurrence of a small but
nonzero cosmological {\em constant}.}  Looking further into the hesitation behavior
would be still useful to acquire more insight into the
time-(non)variability of various coupling constants, probably a related issue which seems to deserve further scrutiny [\cite{fon}].

\bigskip\bigskip
\bcent
{\Large\bf Acknowledgements}
\ecent

I wish to thank Akira Tomimatsu and Kei-ichi Maeda for enlightning discussions on the cosmological solutions, and Koichi Yazaki for those  on the QCD model of nucleons.  I would like to thank Ephraim Fischbach for his valuable comments on the present status of the fifth force.  
I am also indebted to Shinsaku Kitakado and Yoshimitsu Matsui for many stimulating conversations.

\bigskip
\bigskip
\bcent
{\Large\bf Appendices}
\ecent
\renewcommand{\theequation}{\Alph
{section}.\arabic{equation}}
\setcounter{equation}{0}
\appendix

\section{Another attractor solution}

It seems interesting to see what the fate of the matter energy $\rho_{*r}$ would be if $\zeta$ violates \reflef{cfr1-25z}).  Rather unexpectedly, $\rho_{*r}<0$ is evaded automatically.  The solutions for $\zeta$ smaller than the critical value given by \reflef{cfr1-25z}) tend to
the ``vacuum" solution given by ($n=0$ for simplicity)
\beqa
a(t_{*})&=&t^{\alpha_{*}}\quad\mbox{with}\quad
\alpha_{*}=\frac{1}{8}\zeta^{-2}, \label{cfrpa_1}\\
\sigma(t_{*})&=& \frac{1}{2\zeta }\ln t_{*} +\bar{\sigma},  \label{cfrpa_2}\\
\rho_{*r}(t_{*})&=& 0,    \label{cfrpa_3}
\eeqa
where $\bar{\sigma}$ is given by
\beq
\Lambda e^{-4\zeta \bar{\sigma}}=\frac{3-8\zeta^2}{64\zeta^4}.
\label{cfrpa_4}
\eeq
Notice that these agree with \reflef{cfr1-21})-\reflef{cfr1-24}) with $n=0$ for $\zeta =\zeta_{0}\equiv 1/2$.

The two solutions may be depicted schematically as in Fig. 9.  For $\zeta > \zeta_{0}$, the solution with $\rho_{*r}>0$, represented by Track 1, is an attractor.  Suppose we descend along Track 1.  At $\zeta =\zeta_{0}$, one switches to Track 2 for $\rho_{*r}=0$ instead of yielding negative matter energy.  We in fact have examples to show that the lower half of Track 1 is a repeller.

\begin{figure}[tbh]
\hspace*{3.5cm}
\epsfxsize=7cm
\epsffile{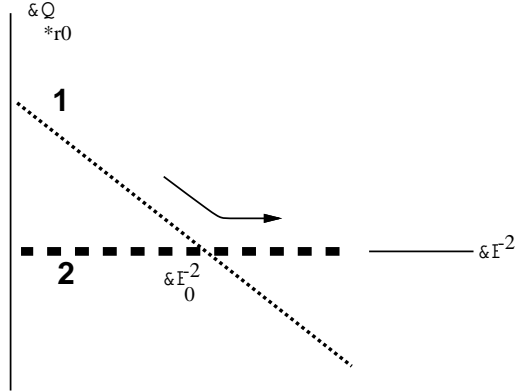}
\caption{The two Tracks 1 and 2 represented by two straight lines
crossing each other correspond to the asymptotic solutions
\protect\reflef{cfr1-23}) and \protect\reflef{cfrpa_3}), respectively.  The horizontal axis is  $\zeta^{-2}$, while the
vertical axis stands for $\rho_{*r0}$.   The upper half of Track 1 and the right half of
Track 2 represent attractors, other portions repellers.  A
``switching'' occurs at the ``crossing'' at $\zeta_{0}^{-2}$. }
\end{figure}

We may respect \reflef{cfr1-25z}) as far as we are interested only in the universe which accommodates nontrivial matter content asymptotically.

\setcounter{equation}{0}
\section{Hesitation behavior}

Suppose there was a period in which $\rho_{*\sigma}\gg\rho_{*r}$ in
the very early universe.  This is reasonbaly expected if reheating
after inflation was not too much sufficient to recover $\rho_{*r}$
which  had been extremely red-shifted during inflation.  Notice that
$\rho_{*\sigma}$ should have 
stayed basically of the order of $\Lambda$ when $\sigma$ was rolling down the slope of the exponential potential as given by \reflef{cfr1-14}).

Also the exponential slope is so steep that $V(\sigma)$ becomes small rapidly, leading to the $K_{\sigma}$-dominated unvierse, where 
\beq
K_{\sigma}=\half \dot{\sigma}^2.
\label{cfrb_1}
\eeq
As is well known only the kinetic energy of a scalar field as the matter density is equivalent to the equation of state $p=\rho$, resulting in the expansion law
\beq
H_{*}=\frac{1}{3}t_{*}^{-1}.
\label{cfrb_2}
\eeq
Using this in \reflef{cfr1-16}) also with  $V'$ ignored we find a solution
\beq
\dot{\sigma}\sim \beta t_{*}^{-1},
\label{cfrb_3}
\eeq
and hence
\beq
\sigma\sim \beta\ln t_{*}.
\label{cfrb_4b}
\eeq
It also follows
\beq
\rho_{*\sigma}\approx K_{\sigma}\sim \half\beta^2 t_{*}^{-2}.
\label{cfrb_4}
\eeq
The coefficent $\beta$ can be determined if we substitute \reflef{cfrb_2}) and \reflef{cfrb_4}) into \reflef{cfr1-15}) also with $V(\sigma)$ dropped:
\beq
\beta =\sqrt{\frac{2}{3}}.
\label{cfrb_4a}
\eeq
In fact the solution given by \reflef{cfrb_2}) and \reflef{cfrb_4b}) with \reflef{cfrb_4a}) is found to be an attractor.

On the other hand, substituting \reflef{cfrb_2}) in \reflef{cfr1-17}) yields
\beq
\rho_{*r}\sim t_{*}^{-4/3},
\label{cfrb_5}
\eeq
which falls off more slowly than $\rho_{*\sigma}$ as given by \reflef{cfrb_4}).  For this reason $\rho_{*r}$ soon catches up and eventually surpasses $\rho_{*\sigma}$.  After this time, the universe enters the radiation-dominated epoch with 
\beq
H_{*}=(1/2)t_{*}^{-1}.  
\label{cfrb_6}
\eeq

Using this in \reflef{cfr1-16}) we obtain
\beq
\dot{\sigma}\sim t_{*}^{-3/2},
\label{cfrb_7}
\eeq
showing that $\sigma$ comes to at rest quickly.  This is the beginning of the hesitation phase.  We also find
\beq
\rho_{*\sigma}\approx K_{\sigma}\sim t_{*}^{-3},
\label{cfrb_8}
\eeq
which decreases much faster than $\rho_{*r}\sim t_{*}^{-2}$.

The potential $V(\sigma) \sim e^{-4\zeta\sigma}$ had already been very small at the onset of the hesitation period, staying there since.  Eventually, however, $K_{\sigma}$ reaches this small value, so that \reflef{cfr1-16}) has to be solved with $V'$ included again, hence the end of the hesitation.

Fig. 5 shows a period in which \reflef{cfrb_2}) is obeyed.  We can also derive the behavior of $a$ in J frame during the $K_{\sigma}$-dominated epoch.  From \reflef{cfrb_4b}) we first find
\beq
\Omega \sim \phi \sim t_{*}^{\zeta\beta}.
\label{cfrb_9}
\eeq
Using this together with \reflef{cfrb_2}) in \reflef{cfr1-20}) we obtain 
\beq
a(t_{*})\sim t^{\tilde{\alpha}}, \quad\mbox{with}\quad
\tilde{\alpha}=\frac{1}{3}-\zeta\beta.
\label{cfrb_10}
\eeq
From \reflef{cfr1-25z}) with $n=0$ and \reflef{cfrb_4a}) we find
\beq
\tilde{\alpha}\leq \frac{1}{3}-\frac{1}{\sqrt{6}}=-0.108 <0,
\label{cfrb_11}
\eeq
implying {\em contraction} of the universe in J frame during the $K_{\sigma}$-dominated universe.  In Fig. 7 we recognize the slope of $\log a$ against $\log t_{*}$ for $\zeta =1$;
\beq
\tilde{\alpha}=\frac{1}{3}-\sqrt{\frac{2}{3}}=-0.483.
\label{cfrb_12}
\eeq

\setcounter{equation}{0}
\section{Loop integrals}

We show some of the details of calculating loop diagrams of Fig. 8.  The basic coupling to the linear $\sigma$ is obtained by expanding  the exponential in \reflef{cfr1-33b2}):
\beq
-\tilde{L}'_{\sigma\Phi}=2\zeta (\nu -2)\sigma\left(
 \half m_{*0}^2\Phi_{*}^2+\frac{1}{4!}\lambda\Phi^4_{*}\right),\label{cfrc_1}
\eeq
where $m_{*0}^2=\xi^{-1}f^2 $.  The diagram (a) comes from the second term of \reflef{cfrc_1}) while the diagram (b) from the first.  The 4-vertex in (b) represents the simple $(\lambda /4!)\Phi^4_{*}$ without $\sigma$ attached.

Considering $\sigma$ simply as a constant, we compute the amplitude for the diagram (a):
\beq
{\cal M}_{a}=-i(2\pi)^{-N}2(\nu -2)\zeta\sigma \lambda \int d^{N}k
\frac{1}{\left( k^2 + m_{*0}^2 \right)}.
\label{cfrc_2}
\eeq
We use
\beq
\int d^{N}k
\frac{1}{\left( k^2 + m_{*0}^2 \right)}=i\pi^2 m_{*0}^2 \Gamma (1-\nu),
\label{cfrc_3}
\eeq
where we have put $N=4$ except inside the $\Gamma$ function.  Substituting \reflef{cfrc_3}) into \reflef{cfrc_2}), and further using
\beqa
(2-\nu )\Gamma(1-\nu )&=&\frac{1}{1-\nu}(2-\nu )\Gamma(2-\nu )
\nnb\\
&=& \frac{1}{1-\nu}\Gamma(3-\nu )\stackrel{\nu \rightarrow 2}
{\longrightarrow} -1,
\label{cfrc_4}
\eeqa
we obtain
\beq
{\cal M}_{a}=\zeta\sigma\frac{\lambda}{8\pi^2}m_{*0}^2.
\label{cfrc_5}
\eeq

Corresponding to (b), we find
\beq
{\cal M}_{b}=i(2\pi)^{-N}2(\nu -2)\zeta\sigma m_{*0}^2\lambda \int d^{N}k
\frac{1}{\left( k^2 + m_{*0}^2 \right)^2 }.
\label{cfrc_6}
\eeq
The integral is now obtained by operating $(-\partial /\partial m_{*0}^2 )$ to the integral in \reflef{cfrc_3}), thus yielding
\beq
{\cal M}_{b}={\cal M}_{a}.
\label{cfrc_7}
\eeq

Adding ${\cal M}_{a}$ and ${\cal M}_{b}$, we finally obtain
\beq
{\cal M}=\zeta\sigma\frac{\lambda}{4\pi^2}m_{*0}^2,
\label{cfrc_8}
\eeq
from which follow \reflef{cfr1-36a}) and \reflef{cfr1-36b}).

\bigskip\bigskip

\begin{center}
{\Large\bf References}
\end{center}
\begin{enumerate}
\item\label{ost}See, for example, J.P. Ostriker and P.J. Steinhardt,
Nature {\bf 377}(1995)600: W.L. Freedman, Measuring
Cosmological Parameters, astro-ph/9706072.
\item\label{dol}A.D. Dolgov, in {\sl The Very Early Universe,} Proc. of Nuffiled Workshop, England, 1982, G.W.Gibbons and S.T. Siklos eds. (Cambridge University Press, Cambridge, Engalnd, 1982): L.H. Ford, Phys. Rev. {\bf D35}(1987), 2339.
\item\label{fn1}Y. Fujii and T. Nishioka, Phys. Rev. {\bf D42}(1990)361, and papers cited therein.
\item\label{bd}C. Brans and R.H. Dicke, Phys. Rev. {\bf
124}(1961)925.

\item\label{hwnt}Y. Fujii and T. Nishioka, Phys. Lett. {\bf B254}, 347(1991): Y. Fujii, Astropart. Phys. {\bf 5}(1996)133: Y. Fujii, in {\sl Proc. of 1st RESCEU International Symposium on ``the Cosmological Constant and the Evolution of the Universe,}" Eds. K. Sato, T. Suginohara and N. Sugiyama, Universal Academy Press, Inc. Tokyo, 1996, p. 155.

\item\label{d1}R.H. Dicke, Phys. Rev. {\bf
125}(1962)2163.
\item\label{wag}See, for example, T. Damour and K. Nordtvedt,
Phys. Rev. {\bf D48}(1993)3436: T. Damour and A.M. Polyakov,
Nucl. Phys. {\bf B423}(1994)532: D.I. Santiago, D. Kalligas and
R.V. Wagoner, Nucleosynthesis Constraints on Scalar-Tensor Theories of 
Gravity, gr-qc/9706017: and papers cited therein.

\item\label{callan}See, for example, C.G. Callan, D. Friedan, E.J. Martinec and M.J. Perry, Nucl. Phys. {\bf B262}(1985)593; C.G. Callan, I.R. Klebanov and M.J. Perry, Nucl. Phys. {\bf B278}(1986)78.
\item\label{halliw}J.J. Halliwell, Phys. Lett. {\bf B185}(1987)341.
\item\label{yokmaed}J. Yokoyama and K.Maeda, Phys. Lett. {\bf B207}(1988)31.
\item\label{barrmaed}J.D. Barrow and K. Maeda, Nucl. Phys. {\bf B341}(1990)294.
\item\label{linde}B.A. Campbell, A. Linde and K.A. Olive, Nucl. Phys. {\bf B355}(1991)146.
\item\label{kol1}S.J. Kolitch and D.M. Eardley, Ann. Phys. {\bf 241}(1995)128.
\item\label{kol2}S.J. Kolitch, Ann. Phys. {\bf 246}(1996)121.
\item\label{santgreg}C. Santos and R. Gregory, Cosmology in Brans-Dicke theory with a scalar potential, gr-qc/9611065.
\item\label{yfq}Y. Fujii, Brans-Dicke cosmology corrected for a
quantum effect due to the scalar-matter coupling, gr-qc/9609044.
\item\label{jd}P. Jordan, {\sl Schwerkraft und Weltalle}, (Friedrich
Vieweg und Sohn, Braunschweig, 1955).
\item\label{fon}Y. Fujii, M. Omote and T. Nishioka, Prog. Theor. Phys. {\bf 92}(1992)521.

\item\label{yf3}Y. Fujii, Mod. Phys. Lett. {\bf A 12}(1997)371.
\item\label{ce}M.S. Chanowitz and J. Ellis, Phys. Ledtt. {\bf 40B}(1972)397.
\item\label{wett}R.D. Peccei, J. Sola and C. Wetterich, Phys. Lett. {\bf B195}(1987)183.
\item\label{loch}M.P.Locher, Proceedings of 12th International Conference on Particle and Nuclei, (MIT) June 1990; Nucl. Phys. {\bf A527} (1991) 73c-88c.
\item\label{fsb}See, for example, E. Fischbach and C. Talmadge, Nature {\bf 356}(1992)207.
\item\label{xj}X. Ji, Phys. Rev. Lett. {\bf 74} (1995) 1071.
\item\label{adel}See, for example, J.H. Gundlach, G.L. Smith, E.G. Adelberger, B.R. Heckel and H.E. Swanson, Phys. Rev. Lett. {\bf 78} (1997) 2523.
\item\label{gdot}See, for example, R.W. Hellings et al, Phys. Rev. Lett. {\bf 51}(1983)1609.
\item\label{saito}A. Kokado, G. Konisi, T. Saito and Y. Tada,Scalar-tensor theory of gravity on $M_{4}\times Z_{2}$ geometry, hep-th/9705195: A. Kokado, G. Konisi, T. Saito and K. Uehara,
Prog. Theor. Phys. {\bf 96}(1996)1291.

\end{enumerate}

\end{document}